\documentclass[sigconf]{acmart}

\usepackage{pifont}

\AtBeginDocument{%
  \providecommand\BibTeX{{%
    \normalfont B\kern-0.5em{\scshape i\kern-0.25em b}\kern-0.8em\TeX}}}

\copyrightyear{2021} 
\acmYear{2021} 
\setcopyright{acmlicensed}\acmConference[AM '21]{Audio Mostly 2021}{September 1--3, 2021}{virtual/Trento, Italy}
\acmBooktitle{Audio Mostly 2021 (AM '21), September 1--3, 2021, virtual/Trento, Italy}
\acmPrice{15.00}
\acmDOI{10.1145/3478384.3478402}
\acmISBN{978-1-4503-8569-5/21/09}

\begin{document}

\title[A New Corpus and Method for Musical Structure Analysis]{A New Corpus for Computational Music Research and\\A Novel Method for Musical Structure Analysis}

\author{Filippo Carnovalini}
\authornote{Funded with a mobility grant from Fondazione Ing. Aldo Gini.}
\email{filippo.carnovalini@dei.unipd.it}
\orcid{0000-0002-2996-9486}
\author{Antonio Rod\`a}
\email{roda@dei.unipd.it}
\affiliation{%
  \institution{Universit\`a degli Studi di Padova}
  \streetaddress{via Gradenigo 6}
  \city{Padova}
  \country{Italy}
}
\author{Nicholas Harley}
\email{nicholas.harley@vub.be}
\author{Steven T. Homer}
\email{sthomer@ai.vub.ac.be}
\affiliation{%
  \institution{Vrije Universiteit Brussel}
  \streetaddress{Pleinlaan 9}
  \city{1050 Brussel}
  \country{Belgium}
}
\author{Geraint A.\ Wiggins}
\email{geraint@ai.vub.ac.be}
\affiliation{%
  \institution{Vrije Universiteit Brussel$^1$ \&\\Queen Mary University of London$^2$}
  \streetaddress{$^1$ Pleinlaan 9; $^2$ Mile End Road, }
\city{$^1$ Pleinlaan 9, 1050 Brussel, Belgium\\$^2$ Mile End Road, London E1 4NS, UK}
  \country{}
}

\renewcommand{\shortauthors}{Carnovalini et al.}

\begin{abstract}
 Computational models of music, while providing good descriptions of melodic development, still cannot fully grasp the general structure comprised of repetitions, transpositions, and reuse of melodic material. 
 We present a corpus of strongly structured baroque allemandes, and describe a top-down approach to abstract the shared structure of their musical content using tree representations produced from pairwise differences between the Schenkerian-inspired analyses of each piece, thereby providing a rich hierarchical description of the corpus.
\end{abstract}

\ccsdesc[500]{Applied computing~Sound and music computing}
\ccsdesc[300]{Information systems~Music retrieval}

\keywords{Computational Musicology, Musical Structure, Music Corpus, Music Generation}


\maketitle

\section{Introduction}
Despite centuries of musicological studies, the advent of computational analysis of music has shed new light on the difficulty of capturing what \textit{defines} a music piece \cite{wiggins_creativity_2020}. This becomes especially evident when computational systems try to generate novel music having learned some features of music from a given dataset of human compositions \cite{briot_deep_2020}. However complex or elegant the model used for the generation, we are still far from obtaining results that are on par with the starting material. This is generally due to the fact that while these models can capture some aspects of the music they analyze, e.g. typical melodic motifs, they fail to capture the entirety of the hierarchical, structural aspects of music. In many cases this leads to algorithms that generate music that sounds reasonable for a short time span, but seems to "wander off" as the length of the generated piece increases \cite{briot_deep_2020-1,carnovalini_open_2019}. 

With this work, we present a small corpus made of twenty-four baroque pieces that exhibit strong structural regularities. This corpus can be used to develop and test algorithms that consider musical structure as a fundamental and primary feature of computational musicology. Though the size of the corpus is relatively small, it is ideal for top-down approaches to musical analysis due to its structural regularity.  This is in contrast to larger, less-structured corpora often used in bottom-up statistical methods, which have resulted in the issues described above.

In the second part of the paper, we also present ongoing research that makes use of the strong regularity of this corpus by using specialized tree representations of musical content to describe musical structure. First, the melody is abstracted through iterated simplifications inspired by Schenkerian analysis, producing what we call \textit{Schenkerian trees} \cite{orio_measure_2009}. Next, the features of these trees are then compared structurally in a pairwise fashion, generating a novel representation that we call \textit{difference trees}. While this is ongoing research and we cannot give a formal evaluation of the results yet, we will show by example how these difference trees can be useful for applications of computational musicology, and discuss further development for computational music generation \cite{carnovalini_computational_2020}. 
\vspace{-0.1cm}
\subsection{Related Works}

This work is linked to a variety of computational musicology applications. Some analysis tools that also abstract tree-like structures based on existing theories of music, such as Schenkerian analysis \cite{marsden_towards_2013,orio_measure_2009} or GTTM \cite{hamanaka_implementing_2016,hamanaka_deepgttm-iii:_2017}, are well known in literature; however, in our proposal the tree representations are not the final goal, but a means for intra-piece comparison in order to analyze the internal repetition structure. The output is similar to other algorithms meant for form analysis \cite{velarde_wavelet-based_2014}, but to our knowledge our approach has never been applied in that field.

Since describing structure is a widely known problem in music generation, some relevant proposals come from that field. GEDMAS \cite{anderson_generative_2013} uses a top-down approach for structured generation, but the melodic content itself is not part of this hierarchical structure. MorpheuS \cite{herremans_morpheus:_2017} applies a structure to imitate a given piece, but there, structure is related to perceived tension, rather than repetition and reuse of melodic content. 
Finally, Wiggins \cite{wiggins_structure_nodate} provides an in-depth theoretical base for the relevance of this approach to music analysis and generation, but does not provide any practical approach to perform the proposed analyses. 

\section{Corpus Description}

The corpus we present is comprised of twenty-four allemandes (dance music originating from Germany, usually possessing even meter), written in 1768 by Gabriele Leone (sometimes referred to as Pietro Leone), a mandolin virtuoso from Naples. These pieces were originally included in a method for teaching mandolin to violin players. As such, despite the great technical ability of the author, the pieces are extremely simple and can be played by a novice. All of the allemandes are written for a mandolin duo, ideally having the first part played by the student and the second by the teacher, so all these pieces are polyphonic. In addition, a single instrument often plays chords, so either part may polyphonic on its own. Since the Neapolitan mandolin has only four strings, neither part ever has more than four simultaneous notes. 
The corpus is released under a Creative Commons license, in MusicXML format.

\subsection{Structure}
Since we propose this corpus as a useful tool to study musical structures, it is worth describing what kind of structural regularities this corpus offers. 

At a high level, there are some evident regularities, as follows:
\begin{itemize}
    \item All of the pieces are divided into two sections, which we call the A section and B section.
    \item In 20 out of 24 pieces, both the A and B sections are eight bars long.
    \item In 22 pieces, the A section is repeated at the end, thus obtaining an "A-B-A" structure.
    \item 20 allemandes have 2/4 meter.
    \item All pieces are in a major key and modulate in the B section either to a close tonality, or in only four of the allemandes, to the minor mode. 
\end{itemize}
The following are the exceptions to above regularities: 
\begin{itemize}
    \item IV has each section repeated (A-A-B-B).
    \item XI has a 12 bars long A section. 
    \item XIV has both A and B section last only 4 bars. 
    \item XVIII has each section repeated and then the A section again (A-A-B-B-A). 
    \item XIX has a 16 bar A section and a 24 bar B section. 
    \item XXI has a 4 bar B section.
    \item VII, IX, XVIII, and XIX have 3/8 meter.
    \item VIII is unique in the corpus in that it has anacrusis.
\end{itemize}
Besides the regularities in the macro-level form, within each of the sections there is frequent use of repetitions, transpositions, and imitation, both within a single part and between parts of the two instruments. These make it simple to distinguish four-measure long phrases that can be further divided into two sub-phrases. 
The relative simplicity of the melodic material makes it easy to distinguish the use of techniques such as repetition, transposition or inversion, making this corpus useful for algorithms that wish to detect such techniques. Moreover, the corpus has two voices, so there is still enough variance and description to extract meaningful information relating to harmony.
One final peculiarity of this corpus not strictly related to musical structure is that each piece was named by the author with an adjective describing the ``feel'' of the piece, like ``The Joyful'', ``The Grumpy'' or ``The Fickle'', and therefore could be used to research if specific musical techniques relate to the proposed emotions and expressions.
Finally, we present the corpus with chord annotations manually added by one of the authors with more than five years of formal music education. 
While there are many MusicXML corpora available, very few present all the above characteristics. For this reason, we believe this corpus can represent a useful tool for many researchers, despite its small size. 

\section{Applications}

In the following subsections, we will describe how the study of this corpus led us to the development of two representations, Schenkerian trees and difference trees, that encode musical structure. 

For the functioning of the following algorithms, the input must first be made of monophonic melodies, with chord annotations that indicate the harmonic development over the melody (lead sheets).
Since we require monophonic melodies, only the first mandolin part was kept, and when two or more notes were played at the same time within the single voice, the higher was chosen. 
The key and meter were annotated as well, but could be inferred with appropriate algorithms if not explicitly present in the input.
In order to compare different moments within a piece, it is necessary to divide the input pieces into smaller segments of equal width and to apply the algorithms below to each segment. A length of one or two bars are reasonable choices, depending on the level of detail that is being considered.

\begin{figure*}
    \centering
    \includegraphics[width=0.9\textwidth]{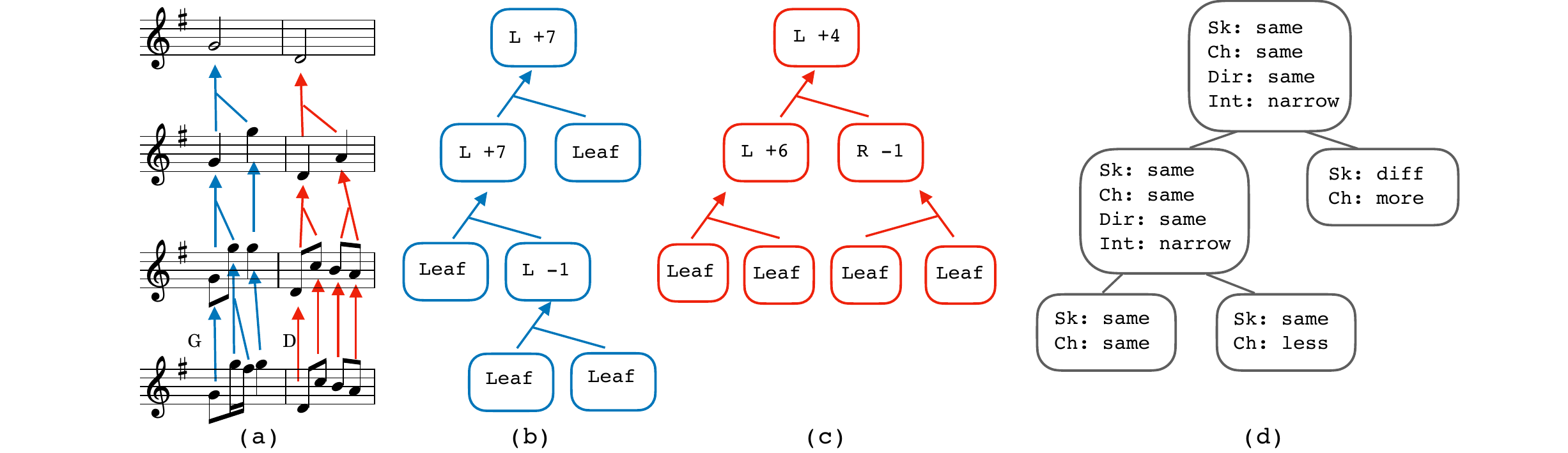}
    \caption{Example of the process building Sk\_trees and a Diff\_tree. (a) shows two bars of music being reduced with the Schenkerian approach. The arrow shows the notes that are kept in the successive reduction, and the line joining the arrow represents the reduced note. (b) shows the Sk\_tree obtained from the first measure. Notice that the leaves correspond to the notes that are present with that duration in the surface melody, regardless of the level. (c) shows the Sk\_tree obtained from the second measure. Notice that there is one node reporting \texttt{R}, as the reduction kept the note on the right, and not the one on the left as usually occurs. (d) shows the Diff\_tree obtained by comparing (a) and (b). Notice that a leaf in this tree occurs whenever a leaf on either Sk\_tree is found.}
    \label{fig:example}
\end{figure*}

\subsection{Schenkerian Tree - Sk\_tree}

The first algorithm takes a segment of a lead sheet as input, and outputs a Schenkeerian tree (Sk\_tree) that represents a set of iterated reductions of the given piece, inspired by what is traditionally done in Schenkerian analysis \cite{schenker_free_1979} or in the Generative Theory of Tonal Music \cite{lerdahl_generative_1985}. Though the details of this algorithm are described in previous related works \cite{orio_measure_2009,simonetta_symbolic_2018,carnovalini_multilayered_2019}, it is worth quickly describing how the algorithm functions at a high-level.

First, we define a sliding window twice as long as the shortest note duration in the input. That window passes over the melody, and whenever there are two or more notes present in window, the ``more important'' note is selected, given the harmonic context, tonality, and metric position of the notes. For example, in the context of a C major piece, over a G chord, a note G will be considered more important than a note F. Next, a new melody made of all the notes that were selected in the first pass is created by extending their duration over the window. For instance, if the first of two quarter notes was selected in the first pass, that note would be extended to a half note in the new melody, and the second note is eliminated.  The size of the window is then increased to twice the shortest duration of this new melody, and the process is iterated until only one note remains. 

These iterated reductions naturally form a tree (see Figure \ref{fig:example}), which we call a Schenkerian tree (Sk\_tree), where the nodes of the tree correspond to the notes of the melody, and each level of the tree represents a level of reduction. The children of a node correspond to the notes present in the window being reduced, with the parent being the result of that reduction.  This note tree is then represented more compactly by annotating how each note is expanded in the lower layer.

\subsection{Difference Tree - Diff\_tree}

The above algorithm serves as a way to simplify the melodic material in piece to make it easier to find regularities, but does not actually compare different segments of a piece to find such regularities. The Difference Trees (diff\_trees) serve this purpose: they compare the reductions made in the Sk\_trees and annotate the actions required to transform from the first tree to the second. The comparisons can be made between any segment of the original piece, but should only be made going forward, i.e. comparing one segment only with segments that come after that, due to the fact that this operation is not commutative. 

The algorithm takes as input two Sk\_trees, and proceeds as follows: the root nodes of the two trees are selected (called R\_1 and R\_2), and a new node is constructed to be the root of the output Diff\_tree. Considering the direct children of R\_1 and R\_2 in their respective trees, in the node of the Diff\_tree the following features are annotated: 
\begin{itemize}
    \item \texttt{Sk}: Schenkerian direction. If R\_1 and R\_2 come from the same position in the children notes (left or right note), annotate \texttt{same}, otherwise \texttt{diff}. A leaf note counts as expanded to the left. 
    \item \texttt{Ch}: Number of children. Annotate if R\_2 has the \texttt{same} number of children as R\_1, or if it has \texttt{more} or \texttt{less}. If the result is same, compare the following features regarding the children:
    \item \texttt{Dir}: Interval direction. Annotate if the interval described in R\_1 has the \texttt{same} direction as the one described by R\_2 or not (\texttt{diff}).
    \item \texttt{Int}: Interval width. Regardless of the direction, annotate if R\_2's interval is more \texttt{narrow}, more \texttt{wide} or the \texttt{same} as R\_1's interval.
\end{itemize}

Then, the first child of both is selected, and the algorithm recursively repeats on all children until there are no more children or if they have a different number of children (a leaf is found), in which case the recursion stops as it is not possible to operate the comparison anymore. 
Figure \ref{fig:example}d shows the result of this process.

\subsection{Musical Analysis}

\begin{figure*}
    \centering
    \includegraphics[trim=0 0 0 10,clip,width=\textwidth]{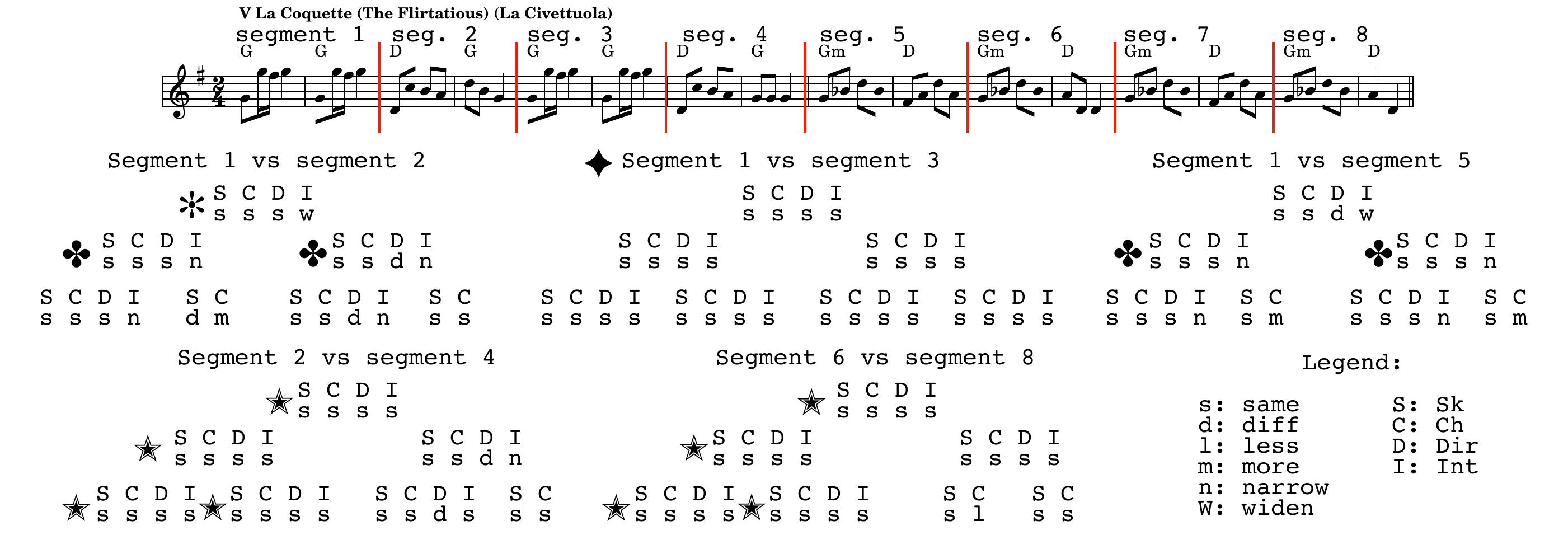}
    \caption{Diffrence trees computed on allemande V, in compact form. The nodes below the third level were omitted.}
    \label{fig:diffs}
\end{figure*}

In this section we provide further examples of trees obtained by applying the above algorithm to the corpus, and describe the insights that can be gathered from this process. 

Figure \ref{fig:diffs} shows some Diff\_trees computed from one allemande taken from the corpus: following are some information that can be inferred by the encoding. In the first tree, comparing segments 1 and 2, the highest level shows an interval that is widened (see \ding{93} in the figure). The highest level of reduction is strongly reliant on the harmonic development. This widening is connected to the fact that the second segment is more harmonically diverse than the first. Conversely, the second level of both that tree and the one comparing segments 1 and 5 (\ding{68}) sees only narrowing intervals. This level is less reliant on the harmony but rather gives an indication of the general melodic contour, and indeed segment 1 (and it's repetition segment 3) is the one with the widest extension in the piece. The mentioned repetition is captured by the tree comparing segments 1 and 3 (\ding{70}), that shows no difference at all across all features. The comparison between segments 2 and 4 and the one between segments 6 and 8 (\ding{77}) also show a repetition (of the first measures of these segments), but also shows the ending variation, that is to be expected from the ending of a phrase/section.

\subsection{Future Directions}
While some insights can be gathered by analyzing a single piece, even more can be found by looking at different pieces together. We are currently developing a way to automatically compare the Diff\_trees obtained by different pieces, possibly reaching a definition of what is the general structure found in a corpus, rather than a specific structure extracted from a piece.
The original motivation behind this project was to study musical representations that could capture medium and long term structures of music to be embedded in a music generation system: the above mentioned general structure extracted from a corpus could define a notion of what is ``typical'' within a certain style, that would be useful within a music generation framework \cite{ritchie_empirical_2007,ritchie_evaluation_2019}.

\section{Conclusions}

In this paper, we presented a corpus of twenty-four baroque allemandes that present structural characteristics useful for those who wish to study automatic ways to analyze structure in music. The original corpus is in the public domain and we release all the additional work done on it under a Creative Commons license to allow and welcome researchers who want to use it as a research tool.
In the second part of the paper, we present the algorithms that we used to analyze such structures, giving an example of how the corpus can be useful to researchers and possibly giving insights for future research developments on structural analysis. We also discussed the further developments that are being done by ourselves.  
While the work presented here is by no means exhaustive of the possible analyses and representations of structure in music, we hope that this contribution can help those who wish to formalize this aspect of music that is fundamental for the creation of meaningful music, but still seems difficult to define in a practical way. 

%

\bibliographystyle{ACM-Reference-Format}
\bibliography{biblio,my_papers}

\appendix
\vspace{-0.1cm}
\section{Online Resources}

The corpus can be accessed at the following link: \\
\href{https://doi.org/10.5281/zenodo.5118650}{https://doi.org/10.5281/zenodo.5118650}

\end{document}